\documentclass[twocolumn]{aastex62}

\shorttitle{}
\shortauthors{Cohen et al.}

\begin{document}

\title{Exoplanet Modulation of Stellar Coronal Radio Emission}

\email{ofer\_cohen@uml.edu}

\author{Ofer Cohen}
\affiliation{Lowell Center for Space Science and Technology, University of Massachusetts Lowell \\
600 Suffolk St., Lowell, MA 01854, USA}

\author{Sofia-Paraskevi Moschou}
\affiliation{Harvard-Smithsonian Center for Astrophysics, 60 Garden St., Cambridge, Massachusetts, USA}

\author{Alex Glocer}
\affiliation{NASA/Goddard Space Flight Center, Greenbelt, Maryland, USA}

\author{Igor V. Sokolov}
\affiliation{Center for Space Environment Modeling, University of Michigan, 2455 Hayward, Ann Arbor, Michigan USA}

\author{Tsevi Mazeh}
\affiliation{School of Physics and Astronomy, Tel Aviv University, Tel Aviv 69978, Israel}

\author{Jeremy J. Drake}
\affiliation{Harvard-Smithsonian Center for Astrophysics, 60 Garden St., Cambridge, Massachusetts, USA}

\author{ C. Garraffo}
\affiliation{Harvard-Smithsonian Center for Astrophysics, 60 Garden St., Cambridge, Massachusetts, USA}

\author{J.~D. Alvarado-G\'omez}
\affiliation{Harvard-Smithsonian Center for Astrophysics, 60 Garden St., Cambridge, Massachusetts, USA}

\begin{abstract}

The search for exoplanets in the radio bands has been focused on detecting radio emissions produced by the interaction between magnetized planets and the stellar wind (auroral emission). Here we introduce a new tool, which is part of our MHD stellar corona model, to predict the ambient coronal radio emission and its modulations induced by a close planet. For simplicity, the present work assumes that the exoplanet is stationary in the frame rotating with the stellar rotation.  We explore the radio flux modulations using a limited parameter space of idealized cases by changing the magnitude of the planetary field, its polarity, the planetary orbital separation, and the strength of the stellar field. We find that the modulations induced by the planet could be significant and observable in the case of hot Jupiter planets ---  above 100\% modulation with respect to the ambient flux in the $10-100~MHz$ range in some cases, and 2-10\% in the frequency bands above $250~MHz$ for some cases. Thus, our work indicates that radio signature of exoplanets might not be limited to low-frequency radio range. We find that the intensity modulations are sensitive to the planetary magnetic field polarity for short-orbit planets, and to the stellar magnetic field strength for all cases. The new radio tool, when applied to real systems, could provide predictions for the frequency range at which the modulations can be observed by current facilities.

\end{abstract}


\section{Introduction} 

The last three decades have provided a prodigious number of exoplanet detections and observations. From knowing very little about exoplanets, even whether they exist at all, this observationally-driven progress has completely revolutionized our understanding of exoplanets and their occurrence in the universe \citep{Deeg17}. The dedicated {\it Kepler} mission \citep{Kepler10} has provided vast statistical information about transiting exoplanet masses, sizes, orbital separations, and inclinations, that is expected to greatly increase with the upcoming {\it TESS} mission \citep{TESS15}. This wealth of transit observations is supplemented by other exoplanet observational techniques, such as radial velocities \citep[see review by][]{Fischer16}, gravitational microlensing \citep[e.g.,][]{Bond04}, and direct imaging \citep[e.g.,][]{Lagrange10}.

The growing amount of new data has led to a shift in the theoretical investigations in the field from {\it detection} to {\it characterization} of the formation, evolution, internal structure, and atmospheres of exoplanets. This theoretical shift is being accompanied by growing observational effort to detect spectral emission from exoplanetary atmospheres \cite[see review][]{Bailey14}, Lyman $\alpha$ signatures of atmospheric evaporation \citep[e.g.,][]{Vidal-Madjar03,Ehrenreich15,Bourrier16,Salz16,Spake18}, and chromospheric signatures of magnetic star-planet interaction \citep[e.g.,][]{shkolnik03,Shkolnik05,Shkolnik08,Fares10,Gurdemir12,Shkolnik18}. Unfortunately, all these characterization methods are extremely hard to realize due to the intrinsically weak planetary signal.  Such data is expected to remain very limited even with the upcoming {\it JWST} mission \citep{JWST06}.  

An additional path in the search and characterization of exoplanets is observations in the radio bands, which can shed light on plasma processes that lead to the generation of radio signal. Relevant processes are expected to operate in the low-frequency range of the radio spectrum, around few tens of MHz and below. However, this range of radio frequencies could be masked by the plasma cutoff frequency of the Earth's ionosphere \citep[e.g.,][]{Davies:69,Yeh.Liu:82},  turning ground-based observations of these radio sources to be extremely challenging. 

Of particular interest in the context of radio observations of exoplanets are the radio waves that are generated by the planet as a radio wave source, in addition to the ambient stellar radio background. Recent theoretical work has been focused on estimating radio emissions that are generated due to the interaction between the stellar wind and the planetary magnetosphere. The interaction leads to particle acceleration that is manifested in auroral emissions and magnetosphere-ionosphere field aligned currents, both associated with known mechanisms to generate radio waves \citep[e.g.,][]{Zarka07,Lazio07,Grismeier07,Vidotto15,See15,Nichols16,Alvarado-Gomez.etal:16b,BurkhartLoeb17,Turnpenney18,Lynch18}.  \cite{See15} have used Zeeman-Doppler imaging (ZDI) maps to estimate the temporal variations of radio emissions from exoplanets directly from the magnetic maps, and using some empirical estimation for the radio power (assuming planetary auroral emissions). \cite{Llama18} presented a more detailed calculation of the coronal radio emissions from V374 Peg using potential field approximation, and hydrostatic coronal density (non-MHD solution). However, both of these studies did not include an actual planet in their simulations. 

Here, we take an alternative approach to investigate the detectability of exoplanets in the radio bands. Instead of detecting the planet {\it as a radio source}, we estimate the {\it planet's induced modulation of the background coronal radio emission}. 
As a starting point, we explore this effect in a limited range of stellar and planetary parameters,
assuming the exoplanet is stationary in the frame rotating with the stellar rotation,
and demonstrate that our model can provide predictions for exoplanetary radio modulations. In particular, we narrow down on the radio bands needed for potential observations of these modulations.

We describe our model and the synthetic radio image tool in Section~\ref{Model}, which is based on the development of \citet{Moschou.etal:18}, and detail the results in Section~\ref{Results}. We discuss our findings and state the next step of our investigation of exoplanetary radio modulations in Sections~\ref{Discussion}, and conclude our work in \ref{Conclusions}.


\section{Synthetic Radio Imaging of Stellar Coronae}
\label{Model}

\subsection{MHD Model for the Stellar Corona}
\label{MHDmodel}

Our method employs a numerical model of the corona and wind of a star computed self-consistently with an orbiting planet stationary in the frame rotating with the stellar rotation. 
To produce a solution for the stellar corona, we use the {\it BATS-R-US} MHD model \citep{Powell:99,Toth.etal:12} and its version for the solar corona and solar wind \citep{Vanderholst.etal:14}. The model is driven by photospheric magnetic field data, while taking into account the stellar radius, mass, and rotation period. The non-ideal MHD equations are solved on a spherical grid which is stretched in the radial direction, taking into account Alfv\'en wave coronal heating and wind acceleration that are manifested as additional momentum and heating terms. The model also takes into account thermodynamic heating and cooling effects, Poynting flux that enters the corona, and empirical turbulent length-scales. We refer the reader to \cite{Sokolov.etal:13} and \cite{Vanderholst.etal:14} for a complete model description and validation. 

The model provides a steady-state, self-consistent, three-dimensional MHD solution for the hot corona and accelerated stellar wind (in the reference frame rotating with the star), assuming the given, data-driven boundary conditions. Thus, it provides the three-dimensional distribution of the plasma density, temperature, velocity and magnetic field (the complete set of MHD plasma properties). With all the plasma parameters defined everywhere, we can also deduce the plasma frequency, $\omega_p$, at each cell, which enables us to track the refraction of the radio waves through the model domain.

In this study, we use a simple, solar-like dipole field of $10~G$, and solar values for the radius, mass, and rotation period. As the baseline for our investigation, we choose to study a Sun-like star, but we also study a case with a stellar dipole field of $100~G$ representing a moderately active star, or non-active M-dwarf star. As we demonstrate in Section~\ref{Results}, this simplified setting might be sufficient to qualitatively capture the main radio modulation effects regardless of the particular star we use. 

The strength of the planetary field is chosen to be $0.3~G$ (Earth-like) and $1~G$ (Jupiter-like), with semi-major axis, $a$, of $6,~9,~12,$ and $15~R_\star$ located along the $x=0$ axis. These distances translate to $0.028,~0.042,~0.056,$ and $0.070~AU$. The planet is embedded as a second boundary condition as described in \cite{Cohen11}, where we use planetary boundary number density value of $10^7$~cm$^{-3}$, and planetary boundary temperature value of $10^4$~K. These values produce a thermal outflow from the planet in the range of $10^6-10^7~g~s^{-1}$, which is much lower than observed in hot jupiters \citep[$10^{10}$~g~s$^{-1}$, e.g.,][]{Vidal-Madjar03,Murray-Clay09,Linsky10}, but is sufficiently high to modulate the background coronal density (a stronger outflow will only intensify the modulations). For reference, the escape rate from Saturn is estimated to be between $10^2-10^4~g~s^{-1}$ \citep{Glocer07}. Future studies that focus on specific planetary systems will require a more detailed planetary and outflow description. Such details could be obtain by coupling the coronal model described here with a model for the planetary magnetosphere \citep[such as code coupling in the Space Weather Modeling Framework, see][]{Toth.etal:12}.

We require at least 10 grid cells across the planetary body in order to properly resolve it well. Therefore, we use grid refinement with very high resolution around the planet so that the grid size near the planet is $\Delta x \le 0.01~R_\star$. In cases where the planet is closer to the star, the initial spherical grid refinement is sufficient. When the planet is further out from the star, we add an additional ring of high resolution along the orbit of the plane. Due to the grid limitations, we set the planet size to be $0.3~R_\star$ which is roughly three Jupiter radii. We performed several tests that have shown that setting a smaller planet size would only require much higher resolution around it, but the results were similar up to $R_p=0.15~R_\star$, and with the radio modulations remain at a similar magnitude. This is because the modulations also occur due to the surrounding plasma around the planet and not only by the planet itself. 

We limit ourselves to the case of steady-state solutions, with a stationary planet, that are viewed from different angles within the orbital plane to mimic the orbital phase. Nevertheless, the simulated radio wave modulations from these static cases can only be enhanced by time-dependent effects due to the extra contributions by the dynamic interaction of the planetary magnetosphere with the stellar corona. Thus, here we provide a lower limit for the radio wave modulations. It is possible that the modulations of the radio waves induced by the planet have a similar magnitude to the stellar variation of the background radio flux, just like a dip in the visible flux might be attributed to a starspot instead of a planet transit. 

\subsection{Synthetic Radio Images}
\label{RadioImages}

A new tool to create synthetic radio images has been recently added to {\it BATS-R-US} \citep{Benkevitch10,Benkevitch12,Moschou.etal:18}. The tool accounts for the free-free Bremsstrahlung radiation that is created in the corona, and propagates through the non-uniform density of the circumstellar medium \citep[e.g.,][]{Kundu:65book,Oberoi.etal:11,Casini.etal:17,Mohan.Oberoi:17}. The wave refraction depends on the local plasma density and the wave frequency. Thus, the radio waves of a given frequency propagate along curved rather than straight lines. The new radio image tool performs ray-tracing of the curved (i.e. not straight Line-of-Sight) propagation of the waves for a particular frequency, and calculates the integrated intensity of the radio wave at the end of the ray path, at a given pixel on the observing plane. The collection of all the pixels provides a radio image for the particular frequency.

The intensity of each pixel, $I_\nu$, for a given frequency, $\nu$, is the integral over the emissivity along the ray. Thus, the intensity is given by
\begin{equation}
I_\nu=\int B_\nu(T) \kappa_\nu ds.
\end{equation}
For Bremsstrahlung emission, where $h\nu<<k_BT$, the Planckian spectral black body intensity is \citep{Karzas.Latter:61}
\begin{equation}
B_\nu(T) = \frac{2k_BT_e\nu^2}{c^2}
\end{equation}
with $k_B$ being the Boltzmann constant, $T_e$ is the electron temperature, and $c$ is the speed of light. The absorption coefficient, $\kappa_\nu$ is
\begin{equation}
\kappa_\nu =\frac{n^2_ee^6}{\nu^2(k_BT_e)^{3/2} m^{3/2}_e c}<g_{ff}>
\end{equation}
Here $n_e$ is the electron number density, $m_e$ is the electron mass, $e$ is the electron charge, and $<g_{ff}>$ is the Gaunt factor, which is assumed to be equal to 10 \citep{Karzas.Latter:61}.

The ray-tracing path for a given angular frequency, $\omega=2\pi \nu$, is defined by the radio wave refraction from one grid cell to the next one. The index of refraction is related to the dielectric permittivity, $\epsilon$, and is given by 
\begin{equation}
n^2=\epsilon=1-\frac{\omega^2_p}{\omega^2},
\end{equation} 
with $\omega^2_p=4\pi e^2 n_e/m_e$ being the plasma frequency (rad~s$^{-1}$). Assuming plasma quasi-neutrality, where the densities of electrons and ions are the same, we can write the mass density as $\rho=m_pn_e$ with $m_p$ being the proton mass. Thus we have
\begin{equation}
\label{Refraction}
\epsilon=1-\frac{\rho}{\rho_c},
\end{equation} 
where $\rho_c=m_pm_e\omega^2/4\pi e^2$ is the critical plasma density at which the refraction index equals zero and the wave cannot be transmitted through. Eq.~\ref{Refraction} demonstrates that higher frequency radio waves can penetrate deeper into the solar atmosphere, where the density is higher. Thus, the synthesized images for higher frequencies capture more detailed structures, such as active regions, in contrast to the lower frequencies synthesized images, that capture the lower density regions at the top of the corona \citep[as demonstrated in][]{Moschou.etal:18}.


\section{Results}
\label{Results}

Our simulations provide the average radio flux intensity over the radio image (for a given frequency and orbital phase). The intensity modulation is estimated by normalizing the flux for each case, frequency, and phase by the associated flux obtained in the case where there is no planet embedded in the simulation (i.e., the ambient flux). It is assumed that in order for the modulations to be observable, the ambient flux itself should be of an observable magnitude (see Section~\ref{Discussion}). 

\begin{figure*}[h!]
\centering
\includegraphics[width=6.in]{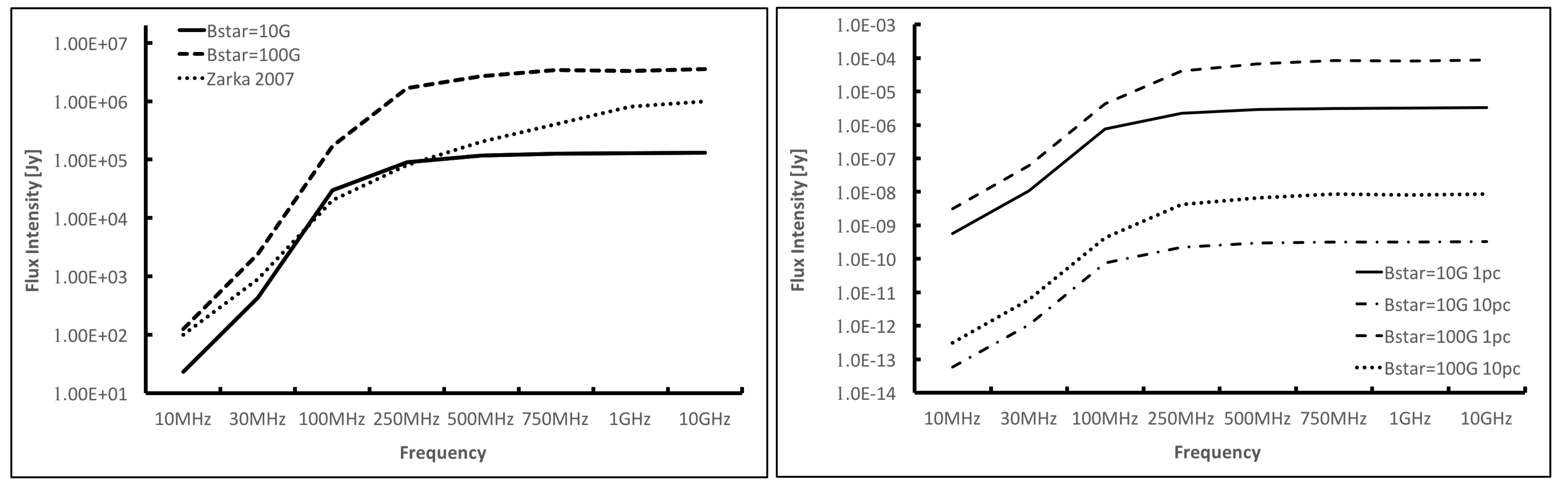}
\caption{Left: synthetic radio flux (in [$Jy$]) as a function of frequency as obtained by our simulations without a planet using dipole stellar magnetic field of $10$ and $100G$. The flux is assumed to be observed from the Earth (from a distance of 1~AU). Also shown the observed radio flux of the quiet Sun taken from \cite{Zarka07}. Right: the same synthetic radio flux spectrum as observed from $1$ and $10~pc$. }
\label{fig:f1}
\end{figure*}

Figure~\ref{fig:f1} shows the radio flux intensity for the different frequencies as obtained from our synthetic radio images for the ambient stellar corona, with stellar magnetic fields of $10~G$ and $100~G$, and without the planet. Overall, the synthetic flux intensities match rather well the observed flux densities for the quiet Sun as observed from the Earth \cite[see Figure 1 in][]{Zarka07}. The deviations are clearly due to the lack of any active regions in our dipolar stellar field, which provide additional radio flux, especially in the higher frequency range. For the same reason, the synthetic radio flux becomes flat above $750~MHz$. The flux in the case of a stellar field of $100~G$ is higher because of an overall increase in the coronal plasma density since the plasma is confined in coronal loops with larger scale-heights, compared to the $10~G$ dipole case. Figure~\ref{fig:f1} also shows how these synthetic radio spectra appear if the source, which is solar-like, is located at
$10~pc$. 

Figure~\ref{fig:f2} shows a three dimensional view of the solutions with planetary magnetic fields of $0.3$ and $1~G$, a stellar dipole field of $10G$, and different orbital separations. The plots are colored with the number density, where selected magnetic field lines are also shown. The star and the planet are shown as red and blue spheres, respectively. The most notable feature in the plots is that the ambient coronal density is modulated by the planet and its magnetosphere. When the planet is closer, at $0.028$ or $0.042~AU$, the lower-density planetary magnetosphere takes over a higher coronal density region, while in the cases of further orbital separations, at $0.056$ and $0.070~AU$, the ambient coronal density is lower than the density of the planetary magnetosphere. It should be noted that the main driver for the modulations in the radio intensity is the density contrast between the ambient corona and the planetary densities at the planetary orbit. This contrast leads to a change in the local plasma frequency, and as a result, the refraction of the radio wave, as well as the radio flux intensity, are modulated. Thus, if the planet significantly modifies the density in a region, it will significantly affect the radio wave refraction, and the question is by how much. Alternatively, if the density contrast between the corona and the planetary magnetosphere is small, we expect weak modulation of the radio flux intensity.

Figure~\ref{fig:f3} and \ref{fig:f4} show a three dimensional view, as well as synthetic radio images for frequencies of $30~MHz$ and $250~MHz$ for cases with planetary field of $0.3~G$ and planet located at $0.028~AU$ and $0.070~AU$, respectively. The plots are displayed from four viewpoints on the simulation domain, which represent four particular phases along the planet's orbit. In the first viewpoint, labeled ``L", the planet appears to the left of the star (pre-transit), in the second viewpoint, labeled ``T", the planet is transiting the star, in the third viewpoint, labeled ``R", the planet appears to the right of the star (post-transit), and in the fourth viewpoint, labeled ``E", the planet is being eclipsed by the star. The local radio flux intensity is in units $[W\;m^{-2}\;Hz^{-1}]$. The figures clearly show that the modulations of the ambient stellar corona plasma by the planet are reflected in the radio images, and that the modulations are different for the different orbital phases. 

Figure~\ref{fig:f5} shows synthetic light curves of the intensity modulation of the radio flux as a function of orbital phase. The transit phase is located in the middle of each plot (at a phase of $0.5$), while the planetary eclipse is located at phase $0$. Each curve represents the relative intensity modulation at a given frequency, with respect to the flux intensity of that frequency for the no-planet, ambient case. The sampled frequencies are $10,~30,~100,~250,~500,~750~MHz$ and $1,~10~GHz$. This range covers the potential frequencies that could be used to detect the planet modulations. It can be seen that in most cases, the low frequency range of $10-30~MHz$ is visibly modulated by the planet. Some weaker modulations occur in the $100~MHz$ and above bands. 

\subsection{Short Orbits Intensity Modulations}

\begin{figure*}[h!]
\centering
\includegraphics[width=5.in]{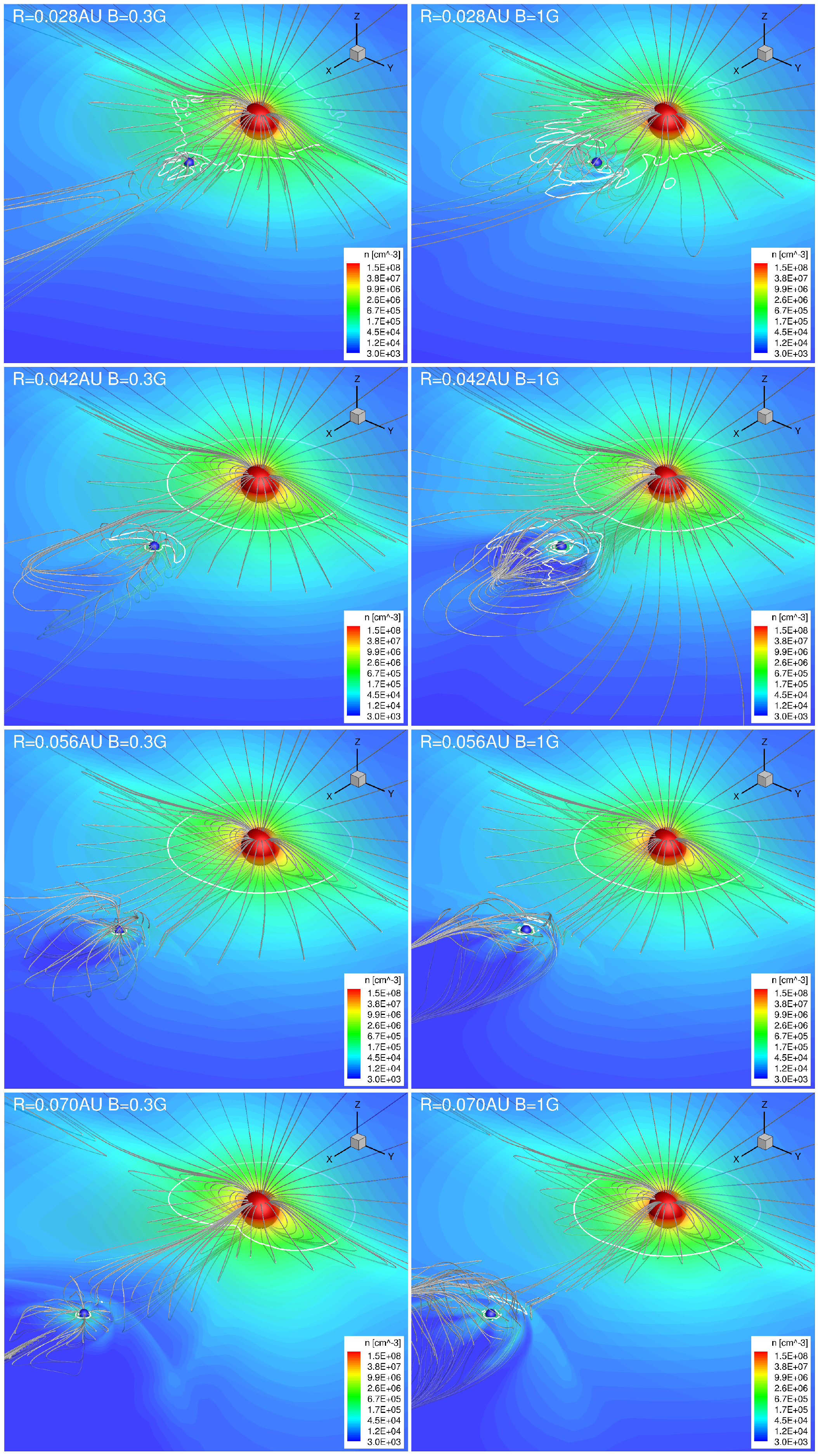}
\caption{Three-dimensional display of the results for planetary field of $0.3G$ (left) and $1G$ (right), for the different orbital separations (top to bottom). Each plot shows the star and planet as red and blue spheres, respectively, color contours of the number density, and selected field lines. The white solid line marks the Alfv\'en surface of the star and the planet.}
\label{fig:f2}
\end{figure*}

\begin{figure*}[h!]
\centering
\includegraphics[width=6.in]{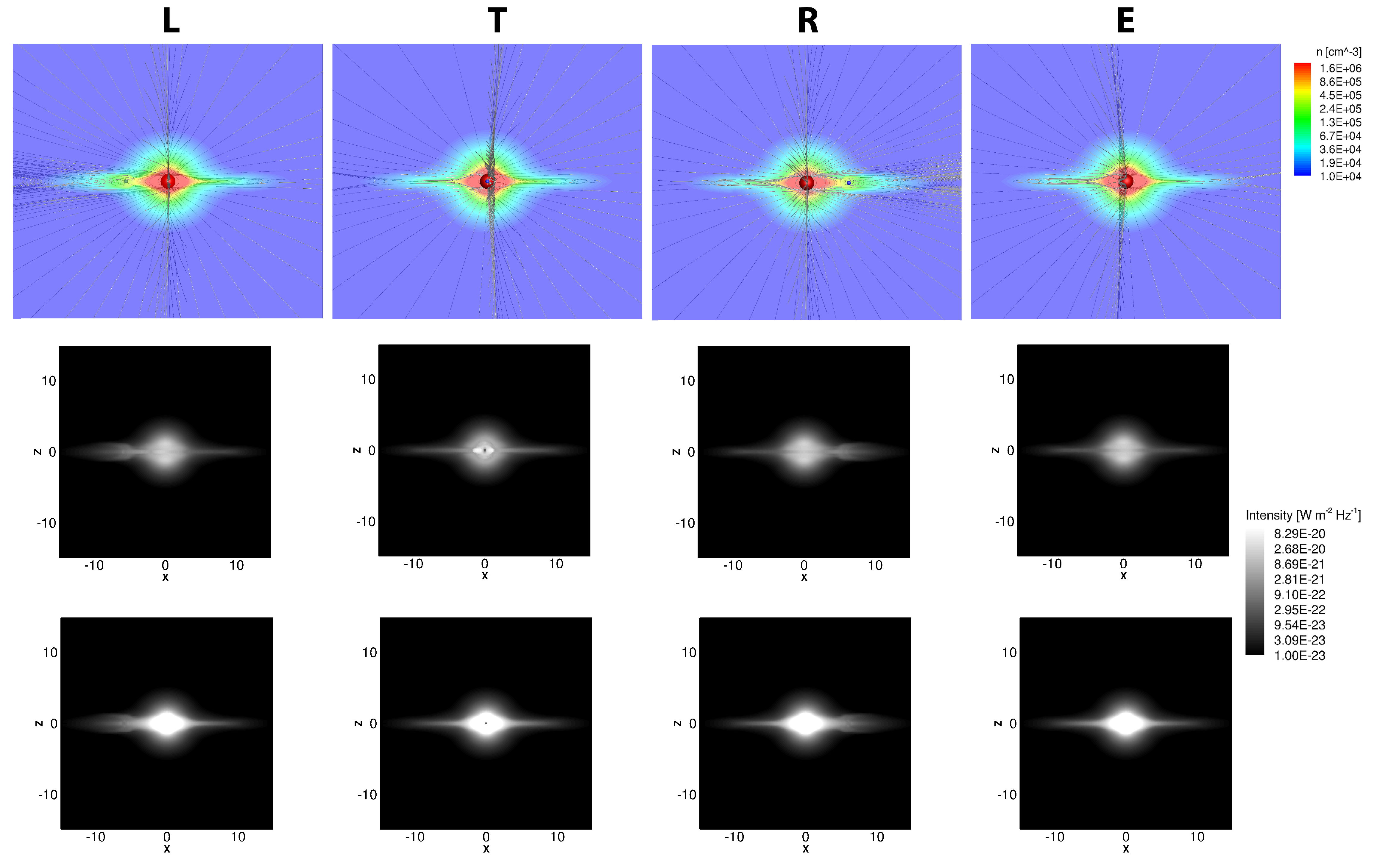}
\caption{Top - similar display as in Figure~\ref{fig:f2} from the different viewing angles on the results for the planet located at $0.028~AU$ and a planetary field of $0.3~G$. The other two rows show the synthetic radio images for $30MHz$ (middle) and $250MHz$ (bottom) for the corresponding viewing angle. The local radio intensity is in units of $W\;m^{-2}\;Hz^{-1}$.}
\label{fig:f3}
\end{figure*}

\begin{figure*}[h!]
\centering
\includegraphics[width=6.in]{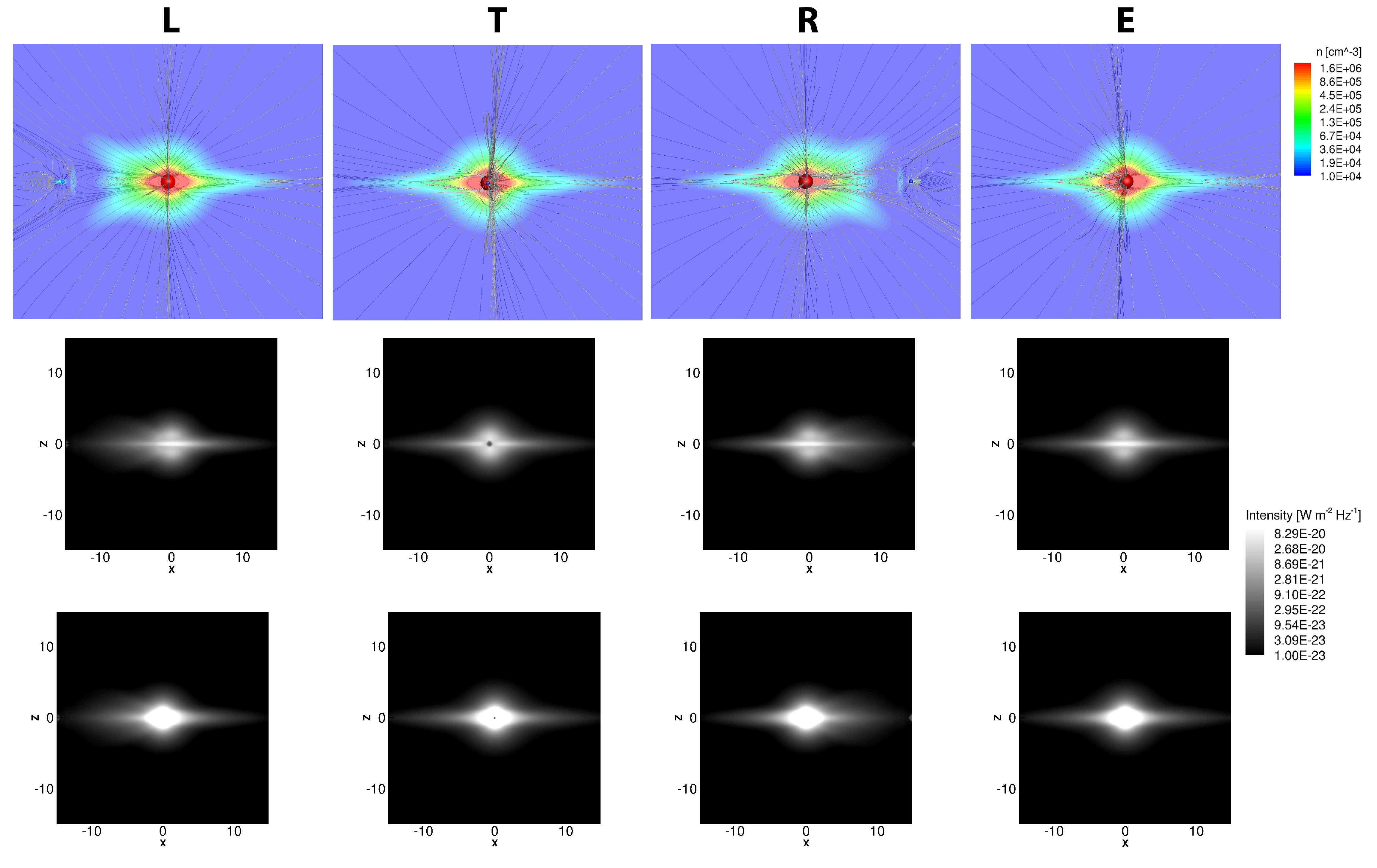}
\caption{Same as Figure=~\ref{fig:f3}, but for the case where the planet is located at $0.070~AU$.}
\label{fig:f4}
\end{figure*}

\begin{figure*}[h!]
\centering
\includegraphics[width=5.5in]{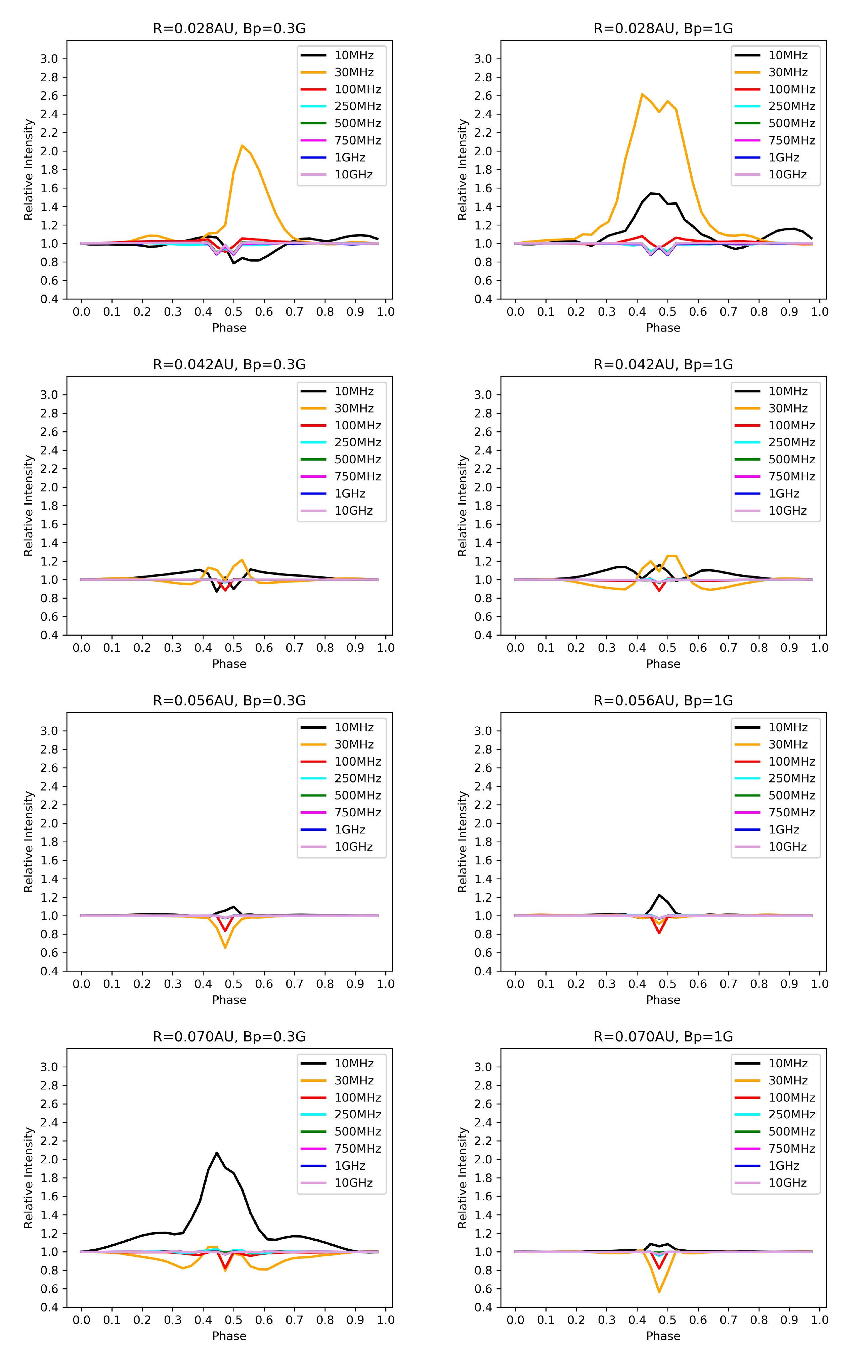}
\caption{Synthetic light-curves of the different frequencies (shown in different colors) for the different test cases. The stellar magnetic field in all cases here is $10~G$. Light-curves show the relative change of intensity with respect to the ambient intensity in the particular band.}
\label{fig:f5}
\end{figure*}

When the planet resides at $a=0.028~AU$, the intensity modulations are driven by the strong star-planet interaction since the planet is located at or close to the Alfv\'en surface (see top panels in Figure~\ref{fig:f2}). While the majority of the background stellar radio emission comes from the dense, helmet streamer regions that face the observer, in the case of $a=0.028~AU$, the edge of the helmet streamer is disrupted by the interaction with the planet. This disruption could involve plasma compression, mixing of coronal and magnetospheric plasmas, as well as creating plasma cavities. As a result, the radio intensity in different bands can be modulated by this local interaction region. Since the helmet streamers emissions are blocked during transit, the contribution to the background emission depends on the emissions generated at the interaction region. 

Looking closely at the synthetic radio images and Figure~\ref{fig:f2}, we find that for the case of planetary field of $1~G$, the intensity contribution of the interaction region in both the $10~MHz$ and $30~MHz$ bands is greater than the ambient intensity from the helmet streamers. Thus, there is a significant intensity increase in these bands during transit. Similar trend is found in the $30~MHz$ intensity for the case with planetary field of $0.3~G$. However, the intensity contribution in the $10~MHz$ with this field strength is found to be negligible comparing to the ambient helmet streamers intensity, which is blocked during transit. Thus, the overall trend we find for this case is an intensity drop in the $10~MHz$ during transit. In the higher frequency bands, even in the $10~GHz$ band, we find a small but noticeable drop of about 10\% in the intensity due to the blockage of the helmet streamers by the planet. 

An interesting feature in the $a=0.028~AU$ and a weak planetary field case is that the emission peaks are slightly shifted from the transit point by about $\sim15$ degrees. This shift seems to happen due to a late but still strong interaction between the planetary magnetosphere and the stellar corona beyond the transit point, and the fact that the interaction between the planetary and stellar plasma occurs at or within the Alfv\'enic point. This a-symmetry is not visible in any of the other cases. 

\subsection{Mid-range Orbits Intensity Modulations}

For the $a=0.042~AU$ cases, we find similar trends to that of the $a=0.028~AU$ case, but with a significantly reduced magnitude, within the 10\% range of intensity increase or decrease. While the planet and the helmet streamers can still interact at this orbit, the interaction is much weaker than the case with $a=0.028~AU$.

For the $a=0.056~AU$ cases, almost no signs of star-planet interaction is visible, with the exception of a small increase in the $10~MHz$ intensity. This increase is slightly larger for the case with a stronger planetary field, where the magnetosphere is larger comparing to the $0.3~G$ planetary field case, and a stronger plasma compression at the magnetopause. Interestingly, the transit shadowing of the helmet streamers emissions in the $100~MHz$ is still visible, with a decrease of almost 20\% in transit. This particular feature, which has significant modulation beyond the very low frequency range, could potentially be observed. 

\subsection{Long Orbits Intensity Modulations}

At larger orbital separation of $a=15~R_\star$ and planetary field of $0.3~G$, there is a significant enhancement in the $10~MHz$ band. This enhancement is due to a cavity created in front of the planet (see bottom-left panel of Figure~\ref{fig:f2}), which seems to compress the top of the helmet streamer and increase the emissions in this band. This cavity does not appear in the case with a stronger planetary field, due to the increase in plasma density near its magnetopause. The intensity of the $30~MHz$ in the case of weaker planetary field is reduce in the form of two "wings" of the light curve. This pattern suggests that the $30~MHz$ band represents the flanks or edge of the planetary magnetosphere, but it is now shadowing the $30~MHz$ ambient emissions. The magnetosphere for the case with planetary field of $1~G$ is larger. As a result, the density of the magnetospheric plasma in this case is slightly lower, leading to a smaller enhancement in the $10~MHz$ band. The ambient emissions in the $30~MHz$ are still shadowed by the planet, but the magnetospheric impact on the shape of the light curve is not noticeable in the case of the stronger planetary field. The transit intensity drop in the $100~MHz$ is still noticeable at this longer orbit.


\section{Discussion}
\label{Discussion}

As shown in recent papers \citep[e.g.,][]{Zarka07,BurkhartLoeb17,Turnpenney18}, the detection of exoplanetary radio emissions as a source of the radio signal (due to auroral emission) seems to be challenging due to the very low flux in the low-frequency range and due to the fact that ground-based observations cannot be made for frequencies below the ionospheric cutoff frequency of $10MHz$. For the ideal Sun-like cases presented here (as seen from Figure~\ref{fig:f1}), the fluxes are obviously too small for detection. However, there are known radio sources, even solar-like stars \citep{Villadsen14}, that are observable and potentially could host a planet. For example, HD 225239, a G2V star, is $18.4~pc$ away from us, and has a radio flux of $0.18~mJy$ in the $8.44~GHz$ band \citep{Wendker95}. In addition, recent radio observations of stars with known exoplanets have revealed feasible radio fluxes. For example, an intensity of  up to few $mJy$ in the $150~MHz$ in HD 189733 and HD 209458 \citep[GMRT,][]{LecavelierDesEtangs11}, and intensity of few $mJy$ in the $1~GHz$ from Proxima Centauri \citep[The Australia Telescope and Anglo-Australian Telescope,][]{Slee03}.

Our results show that close-in exoplanets could modulate the ambient coronal radio emissions by a significant amount (10\% or more). This means that {\it if the ambient flux itself could be observed, so do the modulations} in many cases. The most significant modulations in the stationary cases are seen in the low-frequencies ($10-100~MHz$). This is not surprising, since these frequencies are associated with emissions from coronal regions with lower densities that the planet disturbs the most (higher frequencies are emitted from regions much closer to the stellar surface). From our numerical simulations, we were able to identify two different mechanisms that contribute in the modulation of the stellar radio corona. Our results indicate that the largest modulations of the ambient radio emissions are created by the strong star-planet interaction and the interaction of the planet with the stellar helmet streamers. The other main modulation is created by the shadowing of the emitting streamer regions during transit. This shadowing is visible in a significant manner even in the higher frequency range, and it is probably the main feature that could be observed with current radio observing facilities. 

Our results also indicate that the most notable modulations occur when the planet is very close to the star, and the star-planet interaction is strong, or in the case where the planet is located at rather longer orbits, where it dominates the low-density, ambient plasma, but it still affects the background emission (this effect is probably reduced with greater orbital separation). The modulations are smallest for the intermediate cases, where the ambient plasma is still quite high, but the star-planet interaction is weaker. 

In our investigation, we use a limited parameter space that covers a solar-like stellar magnetic field, two possible planetary fields, and the semi-major axis. In order to extend our parameter space, we also look at the impact of the planetary field polarity, and the magnitude of the stellar field. In both cases, we only test the cases when the planet is very close or far from the star ($6$ and $15~R_\star$), and a planetary field strength of $1G$.

\subsection{The Effect of the Planetary Magnetic Field Polarity}
\label{PMeffect}

\begin{figure*}[h!]
\centering
\includegraphics[width=6.in]{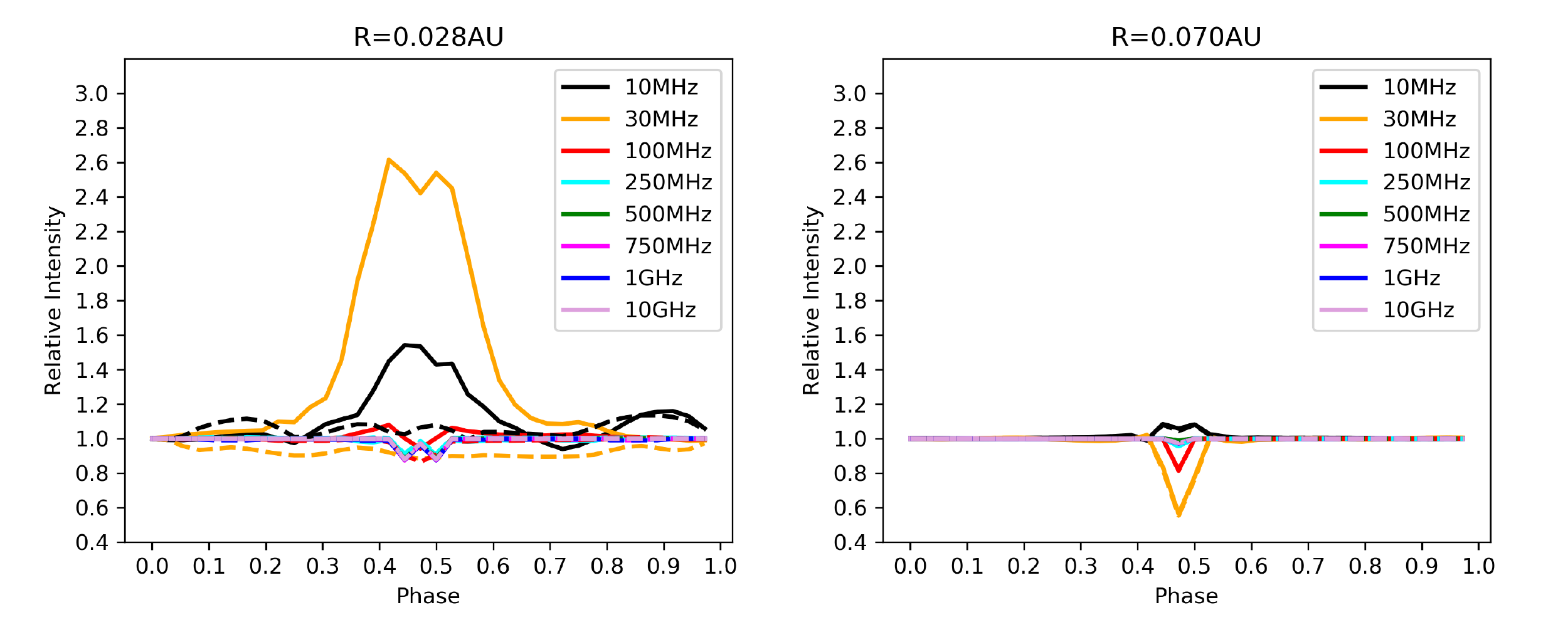}
\caption{Synthetic light-curves of the different frequencies for the case with $a=0.028~AU$ (left) and $a=0.070~AU$ (right), both with a planetary magnetic field strength of $B=1~G$. Solid curves are for planetary field with the same polarity as the stellar field, and dashed curves are for planetary field with polarity opposite to the stellar field polarity. Thus, we expect the modulation to be maximized when the magnetic field polarity of the star and the planet are the same.}
\label{fig:f6}
\end{figure*}

Figure~\ref{fig:f6} compares the modulations for the cases where the planetary field is $\pm1~G$ (with respect to the stellar dipole polarity). It can be clearly seen that when the planet is further away from the star, the results are not affected at all by the polarity of the planetary field. However, when the planet is close to the star, we find a greater difference between the two cases. The planetary field polarity has a stronger impact on the planetary density profile at closer orbits, while the effect is significantly reduced at further orbits. This happens since at closer orbits, where the planet is located at or below the Alfv\'en point, the star-planet interaction is more sensitive to the polarity of the planetary field.  

In particular, the strong enhancements in the $10$ and $30~MHz$ bands for the case where the planetary field polarity is the same as the star are generated by local plasma enhancements through the star-planet interaction. When the field polarity is opposite, magnetospheric plasma is allowed to escape, so the local density enhancements are reduced, resulting in a suppression of the intensity enhancements in the low frequency bands.

\subsection{The Effect of the Stellar Magnetic Field Strength}
\label{StrongB}

\begin{figure*}[h!]
\centering
\includegraphics[width=6.in]{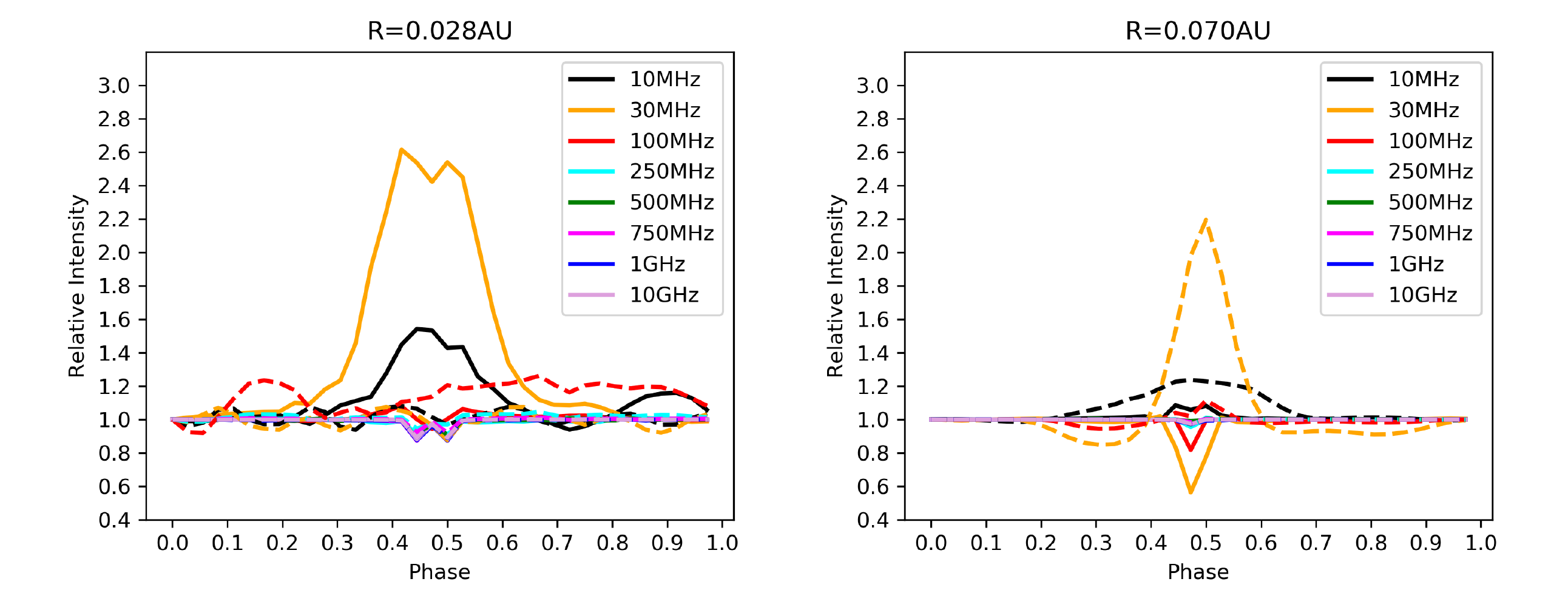}
\caption{Synthetic light-curves of the different frequencies for the case with $a=0.028~AU$ (left) and $a=0.070~AU$ (right), both with a planetary magnetic field strength of $B=1~G$. Solid curves are for stellar dipole field of $10~G$ (same as in Figure~\ref{fig:f2}), and dashed curves are for stellar dipole field of $100~G$. Thus, a stronger stellar magnetic field suppresses the planetary modulation.}
\label{fig:f7}
\end{figure*}

Finally, we investigate how a stronger stellar magnetic field affects the modulations of the coronal radio emission induced by the planet. This is an important factor when considering M-dwarf stars, which are known to potentially have extremely strong magnetic fields of up to few kG \citep{Reiners.Basri:10}, and are much more magnetically active than the Sun. These strong stellar fields could potentially produce strong coronal radio emissions \citep[such as in V374 Peg,][]{Llama18}. We repeat the simulations with a stellar dipole field of $100G$, and the comparison is shown in Figure~\ref{fig:f7}. In general, increasing the stellar dipole strength leads to a larger size of the coronal loops so the helmet streamers extend to a greater distance comparing to the case with a weaker stellar dipole field. The coronal plasma is trapped in these larger closed loops, resulting in an overall enhancement of the coronal density, and a reduction of the density drop with radius. When a planet resides at certain distance from the star, it is surrounded by certain plasma density. By increasing the stellar field strength, we practically move the planet to a higher density region than before. This behavior is clearly seen in Figure~\ref{fig:f7}. 

The case of $a=0.028~AU$ is initially located (with stellar field of $10~G$) at the top of the helmet streamers, and experiences strong interaction with the lower density plasma. This interaction is visible in the $10-30~MHz$ bands. When we increase the stellar field to $100~G$, the planet is surrounded by, and interacts with a much more dense plasma. As a result, the significant modulations in the low frequency range is reduced, and the modulations are more visible in the $100~MHz$ band. When the planet is at $a=0.070~AU$, the increase in the stellar field strength practically move the planet inwards, and the intensity modulation trends resemble the case for weaker stellar field, and a planet at $a=0.028~AU$ (top-right panel of Figure~\ref{fig:f2}).

\subsection{Realistic Radio Observations, Temporal Modulations, and Simulations of Real Planetary Systems}

Our simplified approach here uses a idealized, dipolar stellar magnetic field, and a static, steady-state solutions for the structure of the stellar corona with a planet embedded in it. The phase variations are mimicked by viewing the static, three-dimensional solution from different angles. A number of factors, if included, could immediately provide additional variability of the radio intensity. 

In reality, the structure of the stellar magnetic field and the stellar corona is more complex than the axisymmetric dipolar geometry we use here, and the corona, which hosts the planet, has sectors with different plasma properties along the planetary orbit. Thus, one should expect variations in the plasma density along the planetary orbit as the planet crosses from one plasma sector to another. For short orbits of few days, the size of the coronal sectors should be of the order of the spatial coverage of a large helmet streamer over the orbital plane. Such variations could be captured by modeling more realistic stellar systems. Simulations using ZDI magnetic maps \citep{Donati.etal:89} have been done using our code \citep[e.g.,][]{Cohen.etal:10,Cohen.etal:14,Garraffo.etal:16,Alvarado-Gomez.etal:16a,Alvarado-Gomez.etal:16b,Garraffo17,Pognan18} to simulate the coronae and winds of specific stars. Synthetic radio images could be produced for these more realistic coronal solutions in the same manner of the results presented here. 

The orbital motion of the planet with respect to the ambient coronal plasma, not included in our static simulation, could be included in a time-dependent model as presented in \cite{Cohen11a,Cohen11b}. We expect that the implementation of the planetary orbital motion will enhance the star-planet interaction, and potentially increase the modulations of the radio intensity given the particular geometry of the planetary and stellar magnetic fields (see discussion in Section~\ref{PMeffect}).

An additional factor to consider in a realistic case is the temporal variations of the radio signal itself. Variations in the intensity of the radio signal can be due to photospheric convective and diffusive motions, coronal waves, stellar wind, and stellar rotation. All these processes create temporal density variations in the medium through which the radio waves propagate. These variations extend to a wide range of temporal scales --- seconds, minutes, hours, and days. Of course, it is important to identify these variations in the radio signal in order to isolate the variations that are associated with the planetary orbital motion, which should be of the order of a few hours or more for a planetary orbit of few days. It is also important to remember that radio observations are not the typical time-series of a source, but it is an observation that is quite diffusive in the spatial manner and it is sometimes hard to identify the exact source location due to refraction and scattering effects.

It is reasonable to assume that the short time variations of the order of less than an hour are associated with plasma variations in the low corona and close to the photosphere. Radio emissions of the dense plasma associated with these regions appear in the higher frequency range of the radio spectrum at $1~GHz$ or above. Thus, we expect the ambient large-scale variations of the radio signal, originating from higher altitudes and in the range below $1~GHz$, to have longer time scales than minutes. The coronal helmet streamers tend to stick around for days and even months. Therefore, signal variations in the $10-100~MHz$ range, which originates from the helmet streamers, should have a similar, longer timescale comparing to the variation timescales at higher frequencies. 

Stellar flares can also dramatically disrupt the coronal density structure and affect the radio signal. There is an active search for such radio signal in an attempt to observe stellar Coronal Mass Ejections \citep[CMEs, e.g.,][]{Villadsen17,Crosley18}. Of course, stellar flares can mask the radio modulation created by an exoplanet. However, stellar flares are typically visible in other wavelengths, and a flare-like different emission mechanism would be easy to distinguish by using,  for example, polarization measurements. Therefore, we should have a pretty good idea whether or not a flare occurs during the time of radio exoplanet observation so that period could be excluded to avoid uncertainties. 

Our new radio tool could provide predictions for the frequency range at which it is most likely to detect a signal for specific targets. Such predictions can be used by observational radio facilities, such as LOFAR\footnote{\url{http://www.lofar.org}}, MWA\footnote{\url{http://www.mwatelescope.org/}}, Effelsberg\footnote{\url{https://www.mpifr-bonn.mpg.de/en/effelsberg}}
and VLA\footnote{\url{https://public.nrao.edu/telescopes/vla/}}. It is important to note that here we investigate planets with a very short orbital period to maximize the modulation effect, and we find modulation of 50\% or more in some cases. While we expect that planets with a larger orbital separation will have a much smaller modulation effect, the modulations can still be of the order of few percent. In addition, as described in Section~\ref{StrongB}, in systems where the stellar field is much stronger, such as M-dwarf systems, the reduction due to the orbital separation can be compensated by the increase in stellar field and coronal density.


\section{Conclusions}
\label{Conclusions}

We use the modeling tool presented by \citet{Moschou.etal:18}, which provides synthetic radio images of the free-free Bremsstrahlung stellar coronae radiation, to calculate the modulations of this coronal radio emission by a close-orbit exoplanet. The source of the modulations is the modification of the radio wave refraction pattern as a result of the change in the ambient plasma density by the planet. 

We find that the absolute magnitude of the modulation is significant, and can reach above 100\% in the $10-100MHz$ bands and between 2-10\% in the frequencies above $250MHz$ in some cases. Thus, our model shows that exoplanet radio transit signals could be detectable if the ambient coronal radio emissions are observable, potentially even in the higher-frequency radio bands. We find that the intensity modulation is driven by the star-planet interaction for short-orbit planets, and by the density contrast between the planet and the ambient coronal plasma for longer-orbit planets. We find that the strength of the stellar magnetic field affects the modulation while the polarity of the planetary magnetic field matters only for the short-orbit cases. We plan to apply the new radio tool to specific planetary systems. These simulations will include a realistic stellar magnetic field and the relative motion between the star and the planet.  Thus, such simulations could provide predictions for exoplanetary radio search in these systems.  


\acknowledgments

We thank for an unknown referee for her/his useful comments and suggestions. The work presented here was funded by a NASA Living with a Star grants NNX16AC11G and NASA NExSS grant NNX15AE05G. JDAG was supported by Chandra grants AR4-15000X and GO5-16021X. Simulation results were obtained using the Space Weather Modeling Framework, developed by the Center for Space Environment Modeling, at the University of Michigan with funding support from NASA ESS, NASA ESTO-CT, NSF KDI, and DoD MURI. Simulations were performed on the Massachusetts Green High Performance Computing Center (MGHPCC) cluster.





\end{document}